\documentclass[twocolumn,english]{revtex4-2}
\usepackage[T1]{fontenc}
\usepackage[latin9]{inputenc}
\usepackage{xcolor}
\usepackage{babel}
\usepackage{refstyle}
\usepackage{units}
\usepackage{mathrsfs}
\usepackage{amsmath}
\usepackage{graphicx}
\usepackage{esint}
\PassOptionsToPackage{normalem}{ulem}
\usepackage{ulem}
\usepackage[unicode=true,pdfusetitle,
 bookmarks=true,bookmarksnumbered=false,bookmarksopen=false,
 breaklinks=false,pdfborder={0 0 0},pdfborderstyle={},backref=false,colorlinks=true]
 {hyperref}
\hypersetup{
 citecolor=blue}

\makeatletter

\AtBeginDocument{\providecommand\figref[1]{\ref{fig:#1}}}
\newcommand{\noun}[1]{\textsc{#1}}
\RS@ifundefined{subsecref}
  {\newref{subsec}{name = \RSsectxt}}
  {}
\RS@ifundefined{thmref}
  {\def\RSthmtxt{theorem~}\newref{thm}{name = \RSthmtxt}}
  {}
\RS@ifundefined{lemref}
  {\def\RSlemtxt{lemma~}\newref{lem}{name = \RSlemtxt}}
  {}

\providecolor{lyxadded}{rgb}{0,0,1}
\providecolor{lyxdeleted}{rgb}{1,0,0}
\DeclareRobustCommand{\lyxsout}[1]{\ifx\\#1\else\sout{#1}\fi}

\usepackage{indentfirst}

\makeatother

\begin{document}
\title{Full Self-Consistent Vlasov-Maxwell Solution}
\author{Aur\'elien Cordonnier}
\affiliation{Aix Marseille Univ, Universit\'e de Toulon, CNRS, CPT, Marseille, France}
\author{Guilhem Dif-Pradalier}
\affiliation{CEA, IRFM, F-13108 St. Paul-lez-Durance cedex, France}
\author{Xavier Garbet}
\affiliation{CEA, IRFM, F-13108 St. Paul-lez-Durance cedex, France}
\author{Xavier Leoncini}
\affiliation{Aix Marseille Univ, Universit\'e de Toulon, CNRS, CPT, Marseille, France}
\begin{abstract}
Full self-consistent stationary Vlasov-Maxwell solutions of magnetically
confined plasmas are built for systems with cylindrical symmetries.
The stationary solutions are thermodynamic equilibrium solutions.
These are obtained by computing the equilibrium distribution function
resulting from maximizing the entropy and closing the equations with
source terms that are then computed by using the obtained distribution.
This leads to a self-consistent problem corresponding to solving a
set of two coupled second order non-linear differential equations.
Relevant plasma parameters are introduced and a bifurcation leading
to an improvement of plasma confinement is shown. Conversely in the
improved confinement setting, we exhibit the emergence of a separatrix
in the integrable motion of a charged particles.
\end{abstract}
\maketitle

\section{Introduction}

Within the scope of the search for better containment of Tokamak-based
fusion plasma, understanding the emergence of transport barriers is
a major issue. Indeed, they may give rise to the so-called H-mode
which is in the heart of the current approach of magnetically confined
fusion reactor. And it now seems accepted that the internal transport
barrier (ITB) play a major role in magnetized fusion plasma \cite{Connor2004,Wolf2003}.

When considering magnetized fusion plasma, it is now commonly accepted
that one of the best description of the plasma is the kinetic one
coupled to the Maxwell equations, and due to the low collisionality
of the plasma the Maxwell-Vlasov system becomes de facto the first
choice. The Vlasov equation is a common feature observed when considering
systems with long range interactions, and beyond plasmas a large number
of physical systems are in this category, like for instance gravitational
forces and Coulomb interactions, vortices in two dimensional fluid
mechanics \cite{Onsager49,Edwards74,Weiss91,Chavanis07}, wave-particle
systems relevant to plasma physics \cite{ElskensBook2002,Benisti2015,Benisti07},
Free-Electron Lasers (FELs) \cite{Bonifacio90,Barre04}. In these
settings, long range interacting Hamiltonian systems display some
common dynamical features. Given some initial condition the systems
exhibit a rapid relaxation towards a quasi-stationary state (QSS).
These QSS have extended lifetime, and one way to tackle them in statisitical
mechanics is to use the Lynden-Bell formalism \cite{LyndenBell67,Chavanis-RuffoCCT07}.
In the realm of long range models the Hamiltonian mean field (HMF)
model \cite{Antoni95} has emerged as being a paradigmatic one, displaying
most of the features observed in systems with long range interactions:
non-additivity \cite{Dauxois_book2002}, out of equilibrium phase
transition \cite{Antoniazzi2007,maximum_entropy}, long lived quasi-stationary
states and slow relaxation towards equilibrium. More recently the
Hamiltonian microscopics dynamics has been investigated and displayed
surprising regularity \cite{Bachelard08}, and lead to the understanding
of stationnary states of the problem as a self-consistent infinite
collection of uncoupled thus integrable pendula \cite{Leoncini09b}
and this could be extended to other models \cite{Vandenberg10}.
This regularity lead to the idea that long range systems organized
themselves in terms of self-organized regularity \cite{Leoncini_Chapter2015},
at least when it was possible, i.e the underlying microscopic dynamics
became integrable once an equilibrium was reached and the self-consistent
fields were thus stationary, this was also the case for the true thermodynamic
equilibrium. Given this feature, we will take a similar approach in
the context of plasma physics and following the recent results discussed
in \cite{Laribi2019,Ogawa2019}. We construct complete thermodynamic
equilibrium solutions of the classical Maxwell-Vlasov equations in
a cylindrical geometry in the spirit of already discussed equilibria
\cite{Bennett34,Bennett55,Morse69,Sharma83,Milovanov93}. Indeed
in this geometry the motion of a charged particle in a two-component
magnetic field can be made integrable, while it is not guaranteed
in a toroidal configuration \cite{Cambon2014}. After applying the
recipes we find stationary solutions of the Maxwell-Vlasov problem
in this geometry, are obtained after solving a system of two coupled
nonlinear ordinary differential equations. In this paper, these equations
are computed and then solved and regimes leading to a plasma confinement
are discussed and investigated, the influence of plasma flows is thoroughly
investigated and a bifurcation leading to an improved plasma confinement
is presented; this is reminiscent at least formally to what one expect
from the H-mode. Another important question that arises is whether
or not the integrable individual microscopic dynamics resulting from
the obtained self-consistent field can have a separatrix in their
phase space. Indeed when only taking into account the self-consitency
partially \cite{Laribi2019}, it was shown that no separatrix could
exist, while as will be shown taking into account the full self-consistency
can lead to the emergence of a separatrix. This feature is quite crucial,
as it was shown in \cite{weitzner_nonperiodicity_1999,Cambon2014}
that when moving to a toroidal configuration the breaking of the separatrix
was leading to Hamiltonian chaos and accompanying this large fluctuation
of the magnetic moment are observed and it can not be considered an
adiabatic invariant \cite{Tennyson86,Neishtadt86}, this should affect
the foundations of gyrokinetics \cite{Brizard2007}, and as such
the results obtained through gyrokinetic simulations could not be
considered as ``first principles''.

Finally one last benefit of having exact solutions of the Valsov-Maxwell
system is that besides their physical relevance which can shed some
different point of views on fusion plasma, if proved to be stable,
the solutions could be used as well as tests for numerical codes like
for instance the one discussed in \cite{Grandgirard2006} as part
of their validation process.

The paper is organized as follows, in a first part we briefly recall
the recipes of the problem that were used in \cite{Ogawa2019,Laribi2019}
and extend it by taking into account the poloidal current feedback.
We introduce as well the relevant physical plasma parameters that
are at the core of our analysis. Then in section~\ref{sec:SectionII},
we quickly derive the equations that allows to study the different
possible regimes, the full derivation and computations of a two species
plasma being derived in the appendix. The solutions are studied in
section~\ref{sec:Section 3}, and a bifurcation between two regimes
is exhibited and the individual microscopic dynamics is discussed.
Finally we conclude.

\section{Full self-consistent equilibrium equations\label{sec:SectionII}}

In this section we present the two nonlinear coupled ordinary differential
equations that govern the self-consistent equilibrium solutions of
the Maxwell-Vlasov problem in the considered cylindrical geometry.
In order to do so, let us start by describing our considered setting.

\subsection{Electromagnetic setting}

We consider the problem of an infinite aspect ratio limit of an ideal
Tokamak such that we can consider the torus as a cylinder and the
usual cylindrical coordinates $(r,\:\theta,\:z)$ and associated unit
vectors $(\text{\ensuremath{\mathbf{e}}}_{r},\:\text{\ensuremath{\mathbf{e}}}_{\theta},\:\text{\ensuremath{\mathbf{e}}}_{z})$.
In this setting, we consider a magnetic field with cylindrical symmetry
\textbf{$\text{\ensuremath{\mathbf{B}}}(r)$} in the following form
\begin{equation}
\text{\ensuremath{\mathbf{B}}}=\text{\ensuremath{\mathbf{B}}}_{Plasma}+\text{\ensuremath{\mathbf{B}}}_{Ext}\label{eq:Magnetic_Field_1}
\end{equation}

\noindent where $\text{\ensuremath{\mathbf{B}}}_{Ext}=B_{0}\,\text{\ensuremath{\mathbf{e}}}_{z}$
is an external uniform magnetic field of intensity $B_{0}$ applied
to the plasma, and $\text{\ensuremath{\mathbf{B}}}_{Plasma}$ is the
magnetic field generated by the plasma. In order to comply with the
symmetry we choose to consider magnetic fields that can be expressed
as

\begin{equation}
\text{\ensuremath{\mathbf{B}}}=B_{0}\left[g(r)\,\text{\ensuremath{\mathbf{e}}}_{\theta}+\left(1+k(r)\right)\,\text{\ensuremath{\mathbf{e}}}_{z}\right]\label{eq:Magnetic_Field_2}
\end{equation}

\noindent where $g$ and $k$ are two functions that remain to be
determined, and correspond to the plasma generated field, i.e. $\text{\ensuremath{\mathbf{B}}}_{Plasma}=B_{0}\left(g(r)\,\text{\ensuremath{\mathbf{e}}}_{\theta}+k(r)\,\text{\ensuremath{\mathbf{e}}}_{z}\right)$.
From this we can get an expression of the vector potential 
\begin{equation}
\text{\ensuremath{\mathbf{A}}}=A_{\theta}(r)\,\text{\ensuremath{\mathbf{e}}}_{\theta}+A_{z}(r)\,\text{\ensuremath{\mathbf{e}}}_{z}\label{eq:Potentiel Vecteur 1}
\end{equation}
in a Coulomb gauge which introduces two other related functions $K(r)$
and $G(r)$:

\begin{equation}
A_{\theta}(r)=\frac{B_{0}}{r}\intop_{0}^{r}u\left(1+k(u)\right)\,\mathrm{d}u=\frac{B_{0}}{r}\left(\frac{r^{2}}{2}+K(r)\right)\label{eq:Definition de K}
\end{equation}

\noindent and

\begin{equation}
A_{z}(r)=-B_{0}\intop_{0}^{r}g(u)\,\mathrm{d}u=-B_{0}G(r)\:.\label{eq:Definition de G}
\end{equation}

\noindent So, the magnetic potential $\text{\ensuremath{\mathbf{A}}}(r)$
writes

\begin{equation}
\text{\ensuremath{\mathbf{A}}}(r)=B_{0}\left[\left(\frac{r}{2}+\frac{K(r)}{r}\right)\,\text{\ensuremath{\mathbf{e}}}_{\theta}-G(r)\,\text{\ensuremath{\mathbf{e}}}_{z}\right]\:.\label{eq:Potentiel Vecteur 2}
\end{equation}

Finally, we will assume that there is no electric field, by for instance
considering there is some neutralizing background or that the charge
density is zero using a two species approach (this is detailed in
the appendix \ref{subsec:Appendix A}).

\subsection{Charged particle dynamics}

We will consider the motion of a charged particle in the fields described
previously. We shall assume that we have a classical non-relativistic
point particle with charge $Q=1$ and mass $m=1$. Using the canonical
variables, the motion is Hamiltonian and the Hamiltonian of the system
writes

\begin{equation}
H=\frac{\left(\ensuremath{\mathbf{p}}-\text{\ensuremath{\mathbf{A}}}(\text{\ensuremath{\mathbf{q}}})\right)^{2}}{2}\:,\label{eq:Hamiltonian}
\end{equation}

\noindent where $\text{\ensuremath{\mathbf{p}}}$ and $\text{\ensuremath{\mathbf{q}}}$
form three pairs of canonically conjugate variables.

The associated equations of motions are :

\begin{equation}
\begin{cases}
\begin{array}{c}
\dot{\text{\ensuremath{\mathbf{q}}}}=\\
\dot{\ensuremath{\mathbf{p}}}=
\end{array} & \begin{array}{c}
\ensuremath{\mathbf{p}}-\text{\ensuremath{\mathbf{A}}}\\
\boldsymbol{\nabla}\text{\ensuremath{\mathbf{A}}}\,.\,\left(\ensuremath{\mathbf{p}}-\text{\ensuremath{\mathbf{A}}}\right)
\end{array}\:.\end{cases}\label{eq:Equations du mouvement}
\end{equation}
Given the specific form of the magnetic field and the associated symmetries
(translation along $z$, and rotation around $\theta$), the motion
of charged particles is integrable and we can reduce the system to
an effective one-dimensional Hamiltonian system 
\begin{align}
H & =\frac{1}{2}\left[p_{r}^{2}+\left(\frac{p_{\theta}}{r}-B_{0}\left(\frac{r}{2}+\frac{K(r)}{r}\right)\right)^{2}+\left(p_{z}+B_{0}G(r)\right)^{2}\right]\label{eq:Effective_Hamiltonian}\\
 & =\frac{p_{r}^{2}}{2}+V_{eff}(r)\:,\label{eq:Effective_potential_def}
\end{align}
where $p_{\theta}$ and $p_{z}$ are constants of the motion see for
instance \cite{Laribi2019}.

\subsection{Kinetic approach and equilibrium stationary distribution}

In order to describe the plasma, we take a kinetic point of view and
will consider a one particle distribution function at equilibrium
in order to describe the physical state of the plasma. As mentioned
we consider no electric field and neglect the collisions, so we can
assume that the dynamics of the distribution function is governed
by the Vlasov equation, and in our non-relativistic setting it corresponds
to the conservation of the particle distribution function along the
trajectory of each particle, i.e.:

\begin{equation}
\frac{\mathrm{d}}{\mathrm{d}t}f(\text{\ensuremath{\mathbf{q}}},\text{\ensuremath{\mathbf{p}}},t)=0\label{eq:Vlasov_definition}
\end{equation}

\noindent where $\text{\ensuremath{\mathbf{q}}}$ and $\mathbf{p}$
satisfy (\ref{eq:Equations du mouvement}). More information can be
found in \cite{mihalas_foundations_1999,krall_principles_1986} for
example. The particles are sources for the fields in the Maxwell equations,
and we have the source terms $n$ for the spatial density function
of charges, and $\mathbf{j}$ for the current vector, are given by

\begin{equation}
n(\text{\ensuremath{\mathbf{q}},t})=\intop_{-\infty}^{+\infty}f(\text{\ensuremath{\mathbf{q}}},\text{\ensuremath{\mathbf{p}}},t)\,\mathrm{d}^{3}p\:,\label{eq:Def Densite de charge}
\end{equation}
and

\begin{equation}
\mathbf{j}(\text{\ensuremath{\mathbf{q}}},t)=\intop_{-\infty}^{+\infty}\mathbf{v}f(\text{\ensuremath{\mathbf{q}}},\text{\ensuremath{\mathbf{p}}},t)\,\mathrm{d}^{3}p\:.\label{eq:Def_Densite de courant}
\end{equation}
Since the motion of particles is governed by the magnetic field, this
implies a self-consistent problem \cite{Vlasov38}. In what follows
we derive a possible candidate of the stationary distribution function
by following the steps of the procedure described in \cite{Laribi2019,Ogawa2019}.
Note that the full derivation of the equations is done in the appendix,
and for clarity we decided to go as straight as possible to the self-consistent
equations to be solved.

When looking for a stationary solution of the non self-consistent
Vlasov equation (\ref{eq:Vlasov_definition}), we can rewrite it with
the usual Poisson bracket as

\begin{equation}
\left\{ f,H\right\} =0\:,\label{eq:Vlasov_Stationary}
\end{equation}

\noindent and so any function of $H$ is a solution of the problem.

Furthermore, when building the distribution function coming from integrable
microscopic motion we want to consider the fact that the total energy
of the system $H$, the total momentum along $z$ and the total angular
momentum along $\theta$ are conserved. Accordingly, the Poisson bracket
with one of these conserved quantities is null. So, we can introduce
respectively four Lagrange multipliers $\beta$, $\gamma_{z}$, $\gamma_{\theta}$
and $\gamma_{0}$ in order to impose constraints corresponding to
these conserved quantities, respectively the energy, the momentum
along $z$, the angular momentum and the number of particles conservation.
And, in order to select a solution among the infinite possibilities,
we settled for the one which maximize the entropy

\begin{equation}
S\left[f\right]=-k_{B}\int\limits _{\Gamma}f\ln(f)\,\mathrm{d}\Omega\label{eq:Definition_entropie}
\end{equation}

\noindent where $k_{B}$ is the Boltzmann constant and $\mathrm{d}\Omega$
the infinitesimal volume of the phase space $\Gamma$, with the previously
mentioned constraints. In order to fully characterize our problems,
since we considered solutions with translation invariance, the relevant
quantity is the lineic particle density instead of the total number
of particles which should become infinite in this setting. Another
way to circumvent this is to consider that we have some kind of periodicity
in the cylinder (some kind of flat torus), so that the total $N$
particles are confined within a length $2\pi R$ of the cylinder,
corresponding to a lineic particle density $\lambda=N/2\pi R$ The
solutions to this variational problem are given by a distribution
of the form

\begin{equation}
f\propto e^{-\beta H-\gamma_{z}p_{z}-\gamma_{\theta}p_{\theta}}\:.\label{eq:Expression_de_f}
\end{equation}

\noindent We can get the exact expression knowing the total number
of particles $N$, indeed we choose to normalize $f$ such that
\begin{equation}
N=\int\limits _{\Gamma}f\,\mathrm{d}\Omega\label{eq:Normalisation avec N}
\end{equation}

\noindent and so the proportionality constant is

\begin{equation}
f_{0}=\frac{N}{4\pi^{2}R\left(\frac{2\pi}{\beta}\right)^{\nicefrac{3}{2}}\int\limits _{0}^{+\infty}re^{-ar^{2}-bG(r)-cK(r)-\gamma_{1}}\,\mathrm{d}r}\:,\label{eq:Coefficient de proportionalite de f}
\end{equation}

\noindent with $\gamma_{1}=-\frac{\gamma_{z}^{2}}{2\beta},$$a=\frac{\gamma_{\theta}}{2}\left({\normalcolor B_{0}-\frac{\gamma_{\theta}}{\beta}}\right)$,
$b=-B_{0}\gamma_{z}$, $c=B_{0}\gamma_{\theta}$ . That leads to the
final expression

\begin{equation}
f=\frac{Ne^{-\beta H-\gamma_{z}p_{z}-\gamma_{\theta}p_{\theta}}}{4\pi^{2}R\left(\frac{2\pi}{\beta}\right)^{\nicefrac{3}{2}}\int\limits _{0}^{+\infty}re^{-ar^{2}-bG(r)-cK(r)-\gamma_{1}}\,\mathrm{d}r}\:.\label{eq:Expression de f finale}
\end{equation}
It may be worth noting here that the $\beta$ parameter corresponds
to the thermodynamic temperature 
\begin{equation}
\frac{1}{k_{B}T}=\frac{\delta\mathcal{S}}{\delta\mathsf{\mathcal{E}}}=\beta\:,\label{eq:Equilibrium temperature}
\end{equation}

\noindent and it can be assumed positive. We also insist on the fact
that $\gamma_{\theta}$ and $\gamma_{z}$ are proportional\noun{ }to
the averages of $v_{\theta}=r\dot{\theta}$ and $v_{z}=\dot{z}$ respectively.
In the literature \cite{Ogawa2019}, it has been noted that when
a plasma rotation exists an ITB can exist. Then, it can be expected
that in such states the averages of $v_{\theta}$ and $v_{z}$ are
not null and so are $\gamma_{\theta}$ and $\gamma_{z}$ respectively.

\subsection{Sources of the plasma magnetic field\label{sec:From-the-magnetic}}

Always considering only one species of a charged particle with charge
$Q=1$ and mass $m=1$, we can compute the particle density, and the
current density in the plasma from the form of the resulting distribution
function (\ref{eq:Expression de f finale}) and the Hamiltonian (\ref{eq:Effective_Hamiltonian}),
and extract an explicit form of the source terms which depends on
the functions $G$ and $K$. For instance, the radial density $n$
behaves like

\begin{equation}
n(\text{\ensuremath{\mathbf{q}}})\propto e^{-ar^{2}-bG(r)-cK(r)}\:.\label{eq:Density_explicit}
\end{equation}
We can also compute the proportionality term and express the radial
density $\rho$ given by

\begin{equation}
\rho(r)=\frac{\intop n(\text{\ensuremath{\mathbf{q}}})\,r\,\mathrm{d}\theta\,\mathrm{d}z}{\intop r\,\mathrm{d}\theta\,\mathrm{d}z}=\frac{1}{4\pi rR}\intop n(\text{\ensuremath{\mathbf{q}}})\,r\,\mathrm{d}\theta\,\mathrm{d}z\label{eq:rho(r)_1}
\end{equation}

\noindent as

\begin{equation}
\rho(r)=\frac{1}{\mathscr{V}}e^{-ar^{2}-bG(r)-cK(r)}\label{eq:rho(r)_2}
\end{equation}
with 
\begin{equation}
\mathscr{V}=\frac{4\pi^{2}R\int_{0}^{+\infty}re^{-ar^{2}-bG(r)-cK(r)}\mathrm{d}r}{N}\:.\label{eq:Definition_normalisation_V}
\end{equation}
We notice that as discussed in \cite{Ogawa2019}, the equation (\ref{eq:rho(r)_2})
shows that the equilibrium profile is not flat as soon as $\gamma_{\theta}$
is not zero and it depends on the poloidal magnetic field configuration
when $\gamma_{z}\neq0$. In other words as soon as the plasma moves
the profiles are not flat. Moreover, since we consider an equilibrium
configuration, we obtain as well a non-flat temperature profile but
we have to consider the local radial kinetic temperature profile,
rather than the thermodynamic one (\ref{eq:Equilibrium temperature})
discussed previously. For instance we can compute the average kinetic
energy at a constant radius

\begin{equation}
\varepsilon(\text{\ensuremath{\mathbf{q}}})=\int Hf\,\mathrm{d}^{3}p,\label{eq:Kinetic_energy_density}
\end{equation}
that leads to

\begin{equation}
\varepsilon(\text{\ensuremath{\mathbf{q}}})=\frac{\partial n(\text{\ensuremath{\mathbf{q}}})}{\partial\beta}\label{eq:Kinetic_energy_density_2}
\end{equation}

\noindent which implies that the radial kinetic energy profile is
proportional to the radial density and therefore has the same shape.

\noindent In the same spirit we now compute the source terms of the
plasma magnetic field and move to the current density $\text{\ensuremath{\mathbf{j}}}$.
We start directly from (\ref{eq:Def_Densite de courant}) and the
speeds

\begin{equation}
v_{z}=p_{z}+B_{0}G(r)\label{eq:V_z}
\end{equation}

\noindent and

\begin{equation}
v_{\theta}=\frac{p_{\theta}}{r}-B_{0}\left(\frac{r}{2}+\frac{1}{r}K(r)\right)\:.\label{eq:V_theta}
\end{equation}
So, if we break down $\text{\ensuremath{\mathbf{j}}}$ by component,
the density current along the $\theta$-coordinate is given by
\begin{equation}
j_{\theta}(\text{\ensuremath{\mathbf{q}}})=\intop_{-\infty}^{+\infty}v_{\theta}f\,\mathrm{d}p_{r}\,\mathrm{d}\frac{p_{\theta}\,}{r}\mathrm{d}p_{z}\label{eq:Definition_j_theta}
\end{equation}

\noindent and ends up as

\begin{equation}
j_{\theta}(\text{\ensuremath{\mathbf{q}}})=-\frac{1}{\mathscr{V}}\frac{\gamma_{\theta}}{\beta}re^{-ar^{2}-bG(r)-cK(r)}\:.\label{eq:expression de j_theta}
\end{equation}
For the density current along the $z$-coordinate, we do the same

\begin{equation}
j_{z}(\text{\ensuremath{\mathbf{q}}})=\intop_{-\infty}^{+\infty}v_{z}f\,\mathrm{d}p_{r}\,\mathrm{d}\frac{p_{\theta}\,}{r}\mathrm{d}p_{z}\label{eq:Definition_j_z}
\end{equation}

\noindent and we obtain

\begin{equation}
j_{z}(\text{\ensuremath{\mathbf{q}}})=-\frac{1}{\mathscr{V}}\frac{\gamma_{z}}{\beta}e^{-ar^{2}-bG(r)-cK(r)}\:.\label{eq:Expression de j_z}
\end{equation}
So we finally find
\begin{equation}
\text{\ensuremath{\mathbf{j}}}(\text{r})=-\frac{1}{\mathscr{V}}\left(\frac{\gamma_{\theta}}{\beta}r\,\mathbf{e}_{\theta}+\frac{\gamma_{z}}{\beta}\,\mathbf{e}_{z}\right)e^{ar^{2}-bG(r)-cK(r)}\:,\label{eq:expression de j}
\end{equation}

\noindent or when rewritten as a function of radial density

\noindent 
\begin{equation}
\text{\ensuremath{\mathbf{j}}}(\text{r})=-\left(\frac{\gamma_{\theta}}{\beta}r\,\text{\ensuremath{\mathbf{e}}}_{\theta}+\frac{\gamma_{z}}{\beta}\,\text{\ensuremath{\mathbf{e}}}_{z}\right)\rho(r)\:.\label{eq:j en fonction de la densite}
\end{equation}
Now that the source terms have been computed we may move to the self-consistent
solutions. However we can already notice that the solutions will obey
an interesting condition that is independent of the thermodynamic
temperature:

\begin{equation}
\frac{j_{\theta}(\text{r})}{rj_{z}(\text{r})}=\frac{\gamma_{\theta}}{\gamma_{z}}\:.\label{eq:PropCurrent}
\end{equation}

\subsection{General Self-Consistent Equation\label{sec:General-Self-Consistent-Equation}}

We have computed the currents which depend on the functions $K$ and
$G$ that are defining the vector potential (\ref{eq:Potentiel Vecteur 2})
in Coulomb gauge ($\boldsymbol{\nabla}\cdot\mathbf{A}=0$) which itself
is related to the current through Amp\`ere's law and ends up to be a
Poisson equation
\begin{equation}
\Delta\mathbf{A}=-\mu_{0}\mathbf{\mathbf{j}}\label{eq:Laplacien de A}
\end{equation}

\noindent and so, using the previously computed source terms we obtain
a set of self-consistent equation

\begin{equation}
\begin{cases}
\frac{1}{r}\frac{\partial}{\partial r}\left(\frac{1}{r}\frac{\partial}{\partial r}K(r)\right) & =\kappa_{\theta}e^{-ar^{2}-bG(r)-cK(r)}\\
\frac{1}{r}\frac{\partial}{\partial r}\left(r\frac{\partial}{\partial r}G(r)\right) & =\kappa_{z}e^{-ar^{2}-bG(r)-cK(r)}
\end{cases}\label{eq:self}
\end{equation}

\noindent with

\begin{equation}
\kappa_{\theta/z}=\frac{\mu_{0}}{B_{0}\beta\mathscr{V}}\gamma_{\theta/z}\:.
\end{equation}
A full derivation of these equations when considering a two species
neutral plasma is performed in Appendix~\ref{subsec:Appendix A},
and we end up with the same form as expressions (\ref{eq:self}).

Let us now study more the system (\ref{eq:self}). First in order
to simplify and given the relation (\ref{eq:PropCurrent}), we rescale
the length using a scaling of the type $\widetilde{r}\rightarrow\frac{\gamma_{\theta}}{\gamma_{z}}r$.
Furthermore, if we also do the transformations $\widetilde{G}(r)\rightarrow bG(r)$
and $\widetilde{K}(r)\rightarrow ar^{2}+cK(r)$, and finally we set
$\widetilde{j_{z}}(\widetilde{r})=\alpha e^{-\widetilde{G}(\widetilde{r})-\widetilde{K}(\widetilde{r})}$
where $\alpha=\frac{\left(b\kappa_{z}\right)^{2}}{c\kappa_{\theta}}$
for the current density, we end up with

\begin{equation}
\begin{cases}
\frac{1}{\widetilde{r}}\frac{\partial}{\partial\widetilde{r}}\left(\frac{1}{\widetilde{r}}\frac{\partial}{\partial\widetilde{r}}\widetilde{K}(\widetilde{r})\right) & =\widetilde{j_{z}}(\widetilde{r})\\
\frac{1}{\widetilde{r}}\frac{\partial}{\partial\widetilde{r}}\left(\widetilde{r}\frac{\partial}{\partial\widetilde{r}}\widetilde{G}(\widetilde{r})\right) & =\widetilde{j_{z}}(\widetilde{r})
\end{cases}\:.\label{eq:self(tilde)}
\end{equation}

For convenience we now omit the $\tilde{\,\,}$, and forget the $z$
in $j_{z}$, also since we only have functions depending on $r$,
partial derivatives are simple ones. Working with Eq.~(\ref{eq:self(tilde)})
we have
\begin{equation}
\frac{1}{r}\frac{\mathrm{d}}{\mathrm{d}r}\left(\frac{1}{r}\frac{\mathrm{d}K(r)}{\mathrm{d}r}\right)=\frac{1}{r}\frac{\mathrm{d}}{\mathrm{d}r}\left(r\frac{\mathrm{d}G(r)}{\mathrm{d}r}\right)
\end{equation}
and obtain

\begin{equation}
\frac{\mathrm{d}K(r)}{\mathrm{d}r}=r^{2}\frac{\mathrm{d}G(r)}{\mathrm{d}r}+\alpha_{0}r\label{eq:-49}
\end{equation}
with the integration constant $\alpha_{0}$ that will need to be determined.
Then from the logarithmic derivative of $j(r)$ we obtain

\begin{equation}
\frac{1}{j(r)}\frac{\mathrm{d}j}{\mathrm{d}r}=-\frac{\mathrm{d}G(r)}{\mathrm{d}r}-\frac{\mathrm{d}K(r)}{\mathrm{d}r}\:.\label{eq:-50}
\end{equation}
And by combining these equations and differentiating (\ref{eq:-50})
we end up with
\begin{equation}
\frac{\mathrm{d}^{2}j}{\mathrm{d}r^{2}}=-\left(\frac{2\alpha_{0}}{1+r^{2}}+(1+r^{2})j\right)j+\left(\frac{1}{j}\frac{\mathrm{d}j}{\mathrm{d}r}-\frac{1-r^{2}}{r\left(1+r^{2}\right)}\right)\frac{\mathrm{d}j}{\mathrm{d}r}\:.\label{eq:SelfCurrentTilde}
\end{equation}
So we end up with one second order nonlinear ordinary differential
equation, which once solved gives us the whole properties of the self-consistent
Vlasov-Maxwell stationary state. Before solving it let us first discuss
the conditions that need to be met for typical physical expected conditions,
we insist that in contrast to the analytical work performed \cite{Laribi2019},
here the full self-consistent field is taken into account and possible
moderation effects on the external magnetic field are taken into account,
leading to a set of differential equations (\ref{eq:self}) instead
of just one.

\subsubsection{Constraints and parameters}

Let's take a closer look at the parameters necessary for the integration
of (\ref{eq:SelfCurrentTilde}) in order to construct our Vlasov-Maxwell
stationary solutions. Given the symmetry of the problem, it is natural
to expect that $\frac{dj}{dr}(0)=0$, so only two parameters $\alpha_{0}$
and $j(0)$ are needed to obtain the solution of (\ref{eq:self(tilde)}).
Then after fixing the plasma constants we will have access to the
full Vlasov-Maxwell solution. The problem lies in connecting these
two parameters with the global equilibrium parameters of the plasma
which are $\beta$, $\gamma_{z}$, $\gamma_{\theta},$ and as well
connect these to the external parameters, $B_{0}$ and the lineic
average plasma density $\lambda=N/2\pi R$. Note that we will assume
that the unit length, i.e the typical scale on which particles are
confined, or a typical radius of the cylinder to be equal to $1$,
so $1/R$ has no dimension and can be more considered like for instance
an aspect ratio if we imagine the cylinder as the limit of a torus.
We shall now attempt to compute the two parameters from the global
parameters, and start our analysis with $\alpha_{0}$.

For this purpose let us recall Eq.~(\ref{eq:-49}) and compute the
constant in $r=0$. Tracing back we obtain

\begin{equation}
\alpha_{0}=\left(\frac{\gamma_{z}}{\gamma_{\theta}}\right)^{2}\left[2a+c\left.\left(\frac{1}{r}\frac{\partial K}{\partial r}\right)\right|_{r=0}\right]-b\left.\left(r\frac{\partial G}{\partial r}\right)\right|_{r=0}
\end{equation}
On the one hand,$\left.\left(r\frac{\partial G}{\partial r}\right)\right|_{r=0}=0$
(from (\ref{eq:Definition de G}) we note that $\frac{\partial G(r)}{\partial r}=g(r)$
and $g(r)$ is bounded). On the other hand, from (\ref{eq:Definition de K})
we note that $\left.\left(\frac{1}{r}\frac{\partial K}{\partial r}\right)\right|_{r=0}=k(0)$.
Given our cylindrical geometry, along $z$ and for $r=0$, the magnetic
field $B_{z}(0)$ corresponds to a solenoidal magnetic field which
is the sum of the external $B_{0}$-field and the field due to the
current $\frac{I_{\theta}}{2\pi R}=\int\limits _{0}^{+\infty}j_{\theta}(r)\mathrm{d}r$.
So we end up with

\begin{equation}
\begin{array}{cc}
k(0) & =\frac{\mu_{0}}{B_{0}}\int\limits _{0}^{+\infty}j_{\theta}(r)\mathrm{d}r\\
 & =-\frac{\mu_{0}}{B_{0}}\frac{N}{4\pi^{2}R}\frac{\gamma_{\theta}}{\beta}
\end{array}\:,
\end{equation}
and thus

\begin{equation}
\alpha_{0}=\frac{\gamma_{z}^{2}}{\gamma_{\theta}}\left(B_{0}-\frac{\gamma_{\theta}}{\beta}\right)-\mu_{0}\frac{\gamma_{\theta}^{2}N}{4\pi^{2}R\beta}\:.\label{eq:Valeur_d_alpha_0}
\end{equation}
Regarding $j(0)$, since the vector potential is defined up to some
constants, we end up with 
\begin{equation}
j(0)=\alpha=\frac{\left(b\kappa_{z}\right)^{2}}{c\kappa_{\theta}}=\frac{\gamma_{z}^{4}}{\gamma_{\theta}^{2}}\frac{\mu_{0}}{\beta\mathscr{V}}\:,\label{eq:valeur_de_j0}
\end{equation}
unfortunately $\mathscr{V}$ depends on the integral of the function
$j(r)/j(0)$, $j(r)$ depends on $j(0)$ and the differential equation
(\ref{eq:SelfCurrentTilde}) is nonlinear, so we have some implicit
problem. Fortunately we have as well some constant parameters in $\mathscr{V}$,
that we may adjust. So the strategy in what follows will be to fix
a value of $\alpha_{0}$ and a value of $j(0)$, so we can obtain
the function $j(r)$, from which the equilibrium will be defined.

\section{Solutions\label{sec:Section 3}}

\subsection{Standard equilibrium profiles}

\begin{figure}
\begin{centering}
\includegraphics[width=8cm]{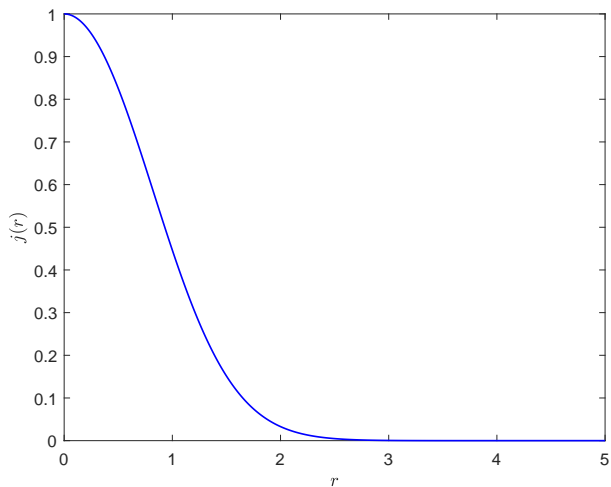}
\par\end{centering}
\caption{\label{fig:Current-density-profile}Typical density profile here obtained
with $j(0)=\alpha_{0}=1$.}
\end{figure}

From the form of the solutions (\ref{eq:SelfCurrentTilde}) and having
the constants (\ref{eq:Valeur_d_alpha_0}) and (\ref{eq:valeur_de_j0})
more or less defined from plasma parameters, we can now compute and
sketch some density current profile. Note that as well we consequently
have access to the density profile since $\rho(r)\propto n(\text{\ensuremath{\mathbf{q}}})\propto j(r)$.
In order to plot these profiles , we have to choose values for the
parameters set $(j(0),\alpha_{0})$. The solutions from the differential
equation (\ref{eq:SelfCurrentTilde}) are computed using octave (lsode)
\cite{Octave}. As we expected from \cite{Laribi2019}, we get as
well non flat ``Gaussian'' type profiles for a given choice of parameters
(see \figref{Current-density-profile}). The quantity $j(0)$, as
we can expect, is linked to the height of the $j(r)$ curve, conversely
$\alpha_{0}$ appears to influence the shape of the profile.

\subsection{Bifurcation towards enhanced confinement profiles}

\begin{figure}
\begin{centering}
\includegraphics[width=8cm]{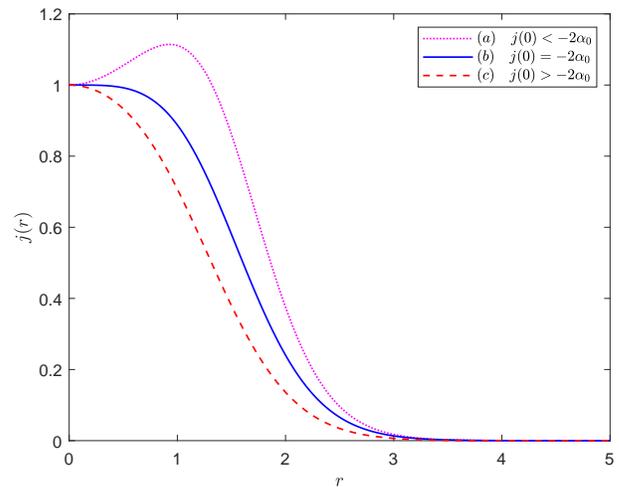}
\par\end{centering}
\caption{\label{fig:Density-profiles_Bifurcation}Density profiles with $j(0)=\alpha=1$.
$(a)$ with $\alpha_{0}=-1$, $(b)$ with $\alpha_{0}=-\nicefrac{1}{2}$,
$(c)$ with $\alpha_{0}=0$. The critical bifurcation value is $\alpha_{0}=-\nicefrac{1}{2}$,
we see an enhanced density profile emerging for $\alpha_{0}<-\nicefrac{1}{2}$.}
\end{figure}

Regarding the behavior of the profile, for a fixed value of $j(0)$
a bifurcation with the emergence of a positive curvature and an enhanced
density profile near $r=0$ can be identified. To do so, we making
some Taylor expansion near $r=0$ and use the self-consistent equation
(\ref{eq:SelfCurrentTilde}). We find that the threshold $\left.\frac{\partial^{2}j}{\partial r^{2}}\right|_{t=0}=0$
is obtained when

\begin{equation}
\frac{j(0)}{-2\alpha_{0}}=1\:.\label{eq:bifurcation_point}
\end{equation}
from which we obtain solutions where the profiles exhibit a maximum
in $r=0$ and others with ``eccentric'' profiles, i.e a maximum
of the density function for a given $r_{0}>0$. In order to study
the different shape of solutions we choose to fix $j(0)=1$ and we
tune the parameter $\alpha_{0}$, results are displayed in \figref{Density-profiles_Bifurcation}.

\begin{figure}
\begin{centering}
\includegraphics[width=8cm]{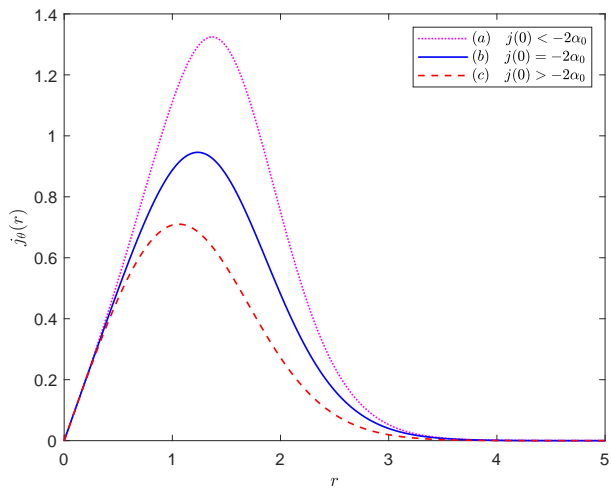}
\par\end{centering}
\caption{\label{fig:-profiles-for_j_theta}$j_{\theta}(r)$ profiles with $j(0)=\alpha=1$.
$(a)$ with $\alpha_{0}=-1$, $(b)$ with $\alpha_{0}=-\nicefrac{3}{2}$,
$(c)$ with $\alpha_{0}=0$. The critical bifurcation value is $\alpha_{0}=-\nicefrac{1}{2}$,
we see that the current profile gets more peaked when $\alpha_{0}<-\nicefrac{1}{2}$.}
\end{figure}

We can notice also the role of the poloidal current density $j_{\theta}(r)$
depicted in \figref{-profiles-for_j_theta}, tends to be stronger
and more peaked, i.e localized, once the bifurcation is crossed. 

\subsection{Link to hyperbolic points}

\begin{figure}
\begin{centering}
\includegraphics[width=8cm]{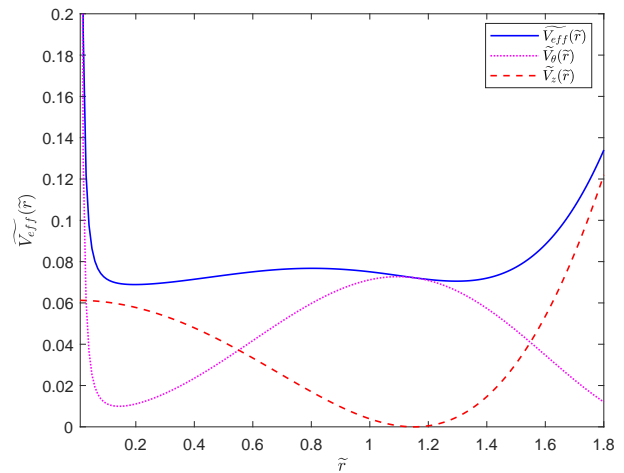}
\par\end{centering}
\caption{Effective potential as a function of $\tilde{r}$, with $j(0)=1$,
$\alpha_{0}=-1$, $p_{\theta}=0.01$ and $p_{z}=0.35$. Both contributions
of the term involving $G$ (dubbed $\tilde{V_{z}}$) and the one involving
$K$ (dubbed $\tilde{V_{\theta}}$) are represented. Both are needed
to explain the shape of the double well potential.\label{fig:Effective_potential_with_unstable_point}}
\end{figure}

Some evidence that steeper density profiles could be linked to the
presence of hyperbolic points in particle trajectories have been made
in \cite{Ogawa2019}. In order to check if this is still the case
with a self-consistent solution let us consider the effective potential
defined in Eq.~(\ref{eq:Effective_potential_def}) and rewrite it
with the scaled variables, we obtain

\begin{equation}
2\gamma_{z}^{2}V_{eff}(r)=\left(\frac{p_{\theta}}{r}-\left(\frac{2\pi\lambda}{\mu_{0}}\frac{r}{2}+\frac{K(r)}{r}\right)\right)^{2}+\left(p_{z}-G(r)\right)^{2}\:.\label{eq:rescaled_effective_hamiltonian}
\end{equation}
To look or hyperbolic points, we need to check the shape of this potential,
which obviously does not depend directly on $\gamma_{z}$, but we
have to choose a value for the lineic density or the ratio $\lambda$
to determine an effective potential. In order to be somewhat realistic,
we settled for an ITER like value of the parameter and fixed $\lambda\sim10^{-20}$,
making the related contribution negligible. We recall that here the
functions $G$ and $K$ are actually $\tilde{G}$ and $\tilde{K}$
and are solutions from the self-consistent equations, and this influenced
by the plasma parameters. Exploring now the shape of the potential
for different values of $p_{\theta}$ and $p_{z}$, we find that there
are effective potentials that give rise to unstable hyperbolic fixed
point (see \figref{Effective_potential_with_unstable_point}), we
find these potentials once we have crossed the bifurcation threshold.

It is important also to point out the influence of diamagnetic effects
due to the poloidal current, indeed when neglecting these effects
it was not possible to obtain effective potentials with hyperbolic
points (see \cite{Laribi2019}). In fact in \figref{Effective_potential_with_unstable_point},
we can see the individual contribution of both terms in the effective
potential, namely the one involving $G$ and the one involving $K$,
and one clearly sees that both are needed to create the hyperbolic
points in between the two two stable elliptic points. Moreover the
presence of such effective potentials above the bifurcation threshold
that creates an enhanced density profile is also consistent with the
results depicted in \cite{Ogawa2019}. This phenomenon could be indeed
important as any perturbation will break the separatrix and lead to
Hamiltonian chaos, like for instance considering these type of magnetic
fields configuration in the torus with large aspect ratios, leads
to chaos and destroys as well the magnetic moment and as such could
impact the reliability of gyrokinetic simulations.

\subsection{Back to plasma parameters}

Now that we have briefly analyzed the solutions that we get, we want
to summarize what are the plasma parameters corresponding to these
solutions and discuss them. We list them in three categories
\begin{description}
\item [{External~constraints}] $R$, $B_{0}$
\item [{Microscopic~physics}] $m^{-}$, $m^{+}$, $q^{-}$, $q^{+}$
\item [{Plasma~parameters}] $N$, $\beta$, $\gamma_{\theta}$, $\gamma_{z}$
\end{description}
We shall below consider only the one species solution discussed previously.
We recall that the characteristic length scale of the systems is given
by 
\begin{equation}
r_{\gamma}=\frac{\gamma_{z}}{\gamma_{\theta}}\:.\label{eq:echelle de longueur}
\end{equation}
We then have $\alpha_{0}$ given by Eq.~(\ref{eq:Valeur_d_alpha_0})
and $j(0)$ by Eq.~(\ref{eq:valeur_de_j0}).

We can then for instance compute the poloidal and toroidal current
by computing the flux of $\boldsymbol{j}$, that leads to the currents

\begin{equation}
I_{\theta}=-\frac{qN}{2\pi}\frac{\gamma_{\theta}}{\beta}\:,
\end{equation}

\begin{equation}
I_{z}=-\frac{qN}{2\pi R}\frac{\gamma_{z}}{\beta}\:.
\end{equation}
Or the typical speed of the plasma along both directions
\begin{equation}
\frac{\left\langle \boldsymbol{v}\right\rangle }{N}=-\left(\begin{array}{c}
0\\
\frac{\gamma_{\theta}}{\beta}\frac{\left\langle r\right\rangle }{N}\\
\frac{\gamma_{z}}{\beta}
\end{array}\right)\:.
\end{equation}
We may as well compute the energy density 
\begin{equation}
\frac{\left\langle H\right\rangle }{N}=\left[\frac{3}{2\beta}+\frac{m\gamma_{\theta}^{2}}{2\beta^{2}}\frac{\left\langle r^{2}\right\rangle }{N}+\frac{m\gamma_{z}^{2}}{2\beta^{2}}\right]\label{eq:densite energie}
\end{equation}
that corresponds to the average kinetic energy of the particles, we
see here that due to the plasma flow we do not have the usual direct
link between $\beta$ and the kinetic energy per particle and additional
terms appear.

In order to see if these stationary solutions could be relevant in
the context of magnetized fusion, we as well compute some order of
magnitudes, considering $T\sim10\text{keV}$, $B_{0}\sim1\,\text{T}$,
$N\sim10^{20}\text{m}^{-3}$, $m\sim10^{-27}\text{kg}$ and $Q=e$.
Let us consider a distribution with $a\sim10$, $b\sim10$ and $c\sim10$,
like what was done in \cite{Ogawa2019}; we also want our typical
scale $r_{\gamma}$ to be of the order of the small radius of a tokamak
so about $1\text{m}$, and some aspect ratio of order $1/3$, this
means $\gamma_{z}\sim\gamma_{\theta}$. With these values, we end
up with $\langle v_{z}\rangle\sim\langle v_{\theta}\rangle\sim c/1000$,
$c$ being the speed of light. And as well $\gamma_{z}\sim\gamma_{\theta}\sim\beta\langle v\rangle\sim5\,10^{-10}\text{USI}$
(corresponding to the international units, note $\gamma_{z}$ and
$\gamma_{\theta}$ do not have the same dimensions but $r\sim1$).
We can as well estimate the current $I_{z}\sim5\,10^{5}\,\text{A}$.
These estimations are in line with typical scales of parameters in
magnetized fusion machines, we may thus anticipate that these stationary
solution could be relevant in the fusion context, and especially the
exhibited bifurcation.

\section{Conclusion and perspectives}

In this paper we have computed a family of stationary solutions of
the Vlasov-Maxwell equations, in a cylindrical geometry. These solutions
correspond to a thermodynamic equilibrium and display a non-uniform
density profile at equilibrium, with as well a non-uniform kinetic
temperature profile, as soon as the plasma displays a collective motion
on the poloidal or ``toroidal'' direction. This simple feature is
already somewhat counter intuitive as the commonly accepted paradigm
in tokamak physics is that these non-uniform profiles are the results
of out of equilibrium features, with energy injection at the center
and dissipation at the walls, so these solutions with global plasma
momentum are offering a possibly different perspective on the confinement.
As shown the solutions are obtained from applying an entropy maximization
principle from which a probability density function is obtained, and
then a self-consistent equation has to be solved on the vector potential
using Maxwell-Amp\`ere equation, that looks like a Poisson equations
and ends up in solving two coupled nonlinear second order ordinary
differential equations. The solutions are described using three intensive
variables $\beta$, $\gamma_{z}$, $\gamma_{\theta}$ corresponding
to the Lagrangian multipliers related respectively to energy, momentum
and angular momentum conservations. From these parameters a typical
scale on which plasma confinement is observed $r_{\gamma}$ emerges
and depends only on the ratio of $\gamma_{z}$ and $\gamma_{\theta}$
, and is as such independent of the global temperature. Moreover,
diamagnetic effects play an important role and a bifurcation between
solutions showing an enhanced confinement profile from more regular
one is displayed and the threshold computed. Finally, when the bifurcation
is crossed and confinement is enhanced, there are regions in phase
space where individual particles are subject to a double well potential
exhibiting a separatrix. The presence of this separatrix in these
enhanced confinement profile is consistent to what was previously
anticipated in a non self-consistent setting \cite{Ogawa2019} and
are as well roots for Hamiltonian chaos under any perturbations, that
can also break the magnetic moment conservation \cite{Cambon2014},
and create some possible problems regarding the validity of gyrokinetic
simulations.

Eventhough computed through a maximizing principle, the stability
of these solutions under for instance a small perturbation like moving
the system to a torus with a large aspect ratio is not at all given.
A perspective of this work would then be to assess the stability of
these solutions, to check also what happens near the separatrices
regarding chaos and the breaking of the magnetic moment, when moving
to a real toroidal geometry and the poloidal symmetry is lost.
\begin{acknowledgments}
This work has been carried out within the framework of the EUROfusion
Consortium, funded by the European Union via the Euratom Research
and Training Programme (Grant Agreement No 101052200 --- EUROfusion).
Views and opinions expressed are however those of the author(s) only
and do not necessarily reflect those of the European Union or the
European Commission. Neither the European Union nor the European Commission
can be held responsible for them.
\end{acknowledgments}

\appendix

\section{Generalization to a two species system with charges $q^{+}$ and
$q^{-}$, and mass $m^{+}$ and $m^{-}$\label{subsec:Appendix A}}

We derive below the full self-consistent system that give rise to
a stationary solution of the Vlasov-Maxwell system. We follow the
same path as the one used for only one species. We use the notation
with a + or a -, at the upper corner, to simplify the notations corresponding
to each species, for example the test particles Hamiltonians write
\begin{equation}
H^{\pm}=\frac{\left(\mathbf{p}^{\pm}-q^{\pm}\mathbf{A}\right)^{2}}{2m^{\pm}}\:,\label{eq:}
\end{equation}
and then lead to the distributions functions

\begin{equation}
f^{\pm}=f_{0}^{\pm}e^{-\beta H^{\pm}-\gamma_{z}^{\pm}p_{z}^{\pm}-\gamma_{\theta}^{\pm}p_{\theta}^{\pm}-\gamma_{1}^{\pm}}
\end{equation}

\noindent after Lagrange multipliers introduction and maximization
of the entropy. Note that by doing so, we assume that the entropy
is additive so the global maximum may be the sum of two maxima taken
for each species individually, which neglect somehow the couplings
through the current for instance, so this may not be an thermodynamic
equilibrium in the end, but anyhow this leads to a stationary solution
of the Vlasov-Maxwell system. We assume that each distribution is
a stationary solution of the Vlasov so that

\begin{equation}
\left\{ f^{\pm},H^{\pm}\right\} =0\:.
\end{equation}

\noindent If we take $a^{\pm}=\frac{\gamma_{\theta}^{\pm}}{2}\left({\normalcolor q^{\pm}B_{0}-\frac{m^{\pm}\gamma_{\theta}^{\pm}}{\beta^{\pm}}}\right)$,
$b^{\pm}=-q^{\pm}B_{0}\gamma_{z}^{\pm}$, $c^{\pm}=q^{\pm}B_{0}\gamma_{\theta}^{\pm}$
and $\gamma_{1}^{\pm}=-\frac{m^{\pm}\left(\gamma_{z}^{\pm}\right)^{2}}{2\beta^{\pm}}$,
the normalization of each distribution function can be derived through

\begin{equation}
\begin{aligned}N^{\pm} & =\int f^{\pm}\,\mathrm{d}^{3}p^{\pm}\mathrm{d}^{3}q^{\pm}\\
 & =f_{0}^{\pm}4\pi^{2}R\left(\frac{2\pi m^{\pm}}{\beta^{\pm}}\right)^{\nicefrac{3}{2}}\int_{0}^{+\infty}re^{-a^{\pm}r^{2}-b^{\pm}G(r)-c^{\pm}K(r)-\gamma_{1}^{\pm}}\mathrm{d}r
\end{aligned}
\end{equation}

\noindent for $N^{\pm}$ the numbers of particles. So the normalization
of $f$ is

\begin{equation}
f_{0}^{\pm}=\frac{N^{\pm}}{4\pi^{2}R\left(\frac{2\pi m^{\pm}}{\beta^{\pm}}\right)^{\nicefrac{3}{2}}\int_{0}^{+\infty}re^{-a^{\pm}r^{2}-b^{\pm}G(r)-c^{\pm}K(r)-\gamma_{1}^{\pm}}\mathrm{d}r}
\end{equation}

\noindent we can then compute the spatial densities for each species

\begin{equation}
\begin{aligned}n^{\pm}(\text{\ensuremath{\mathbf{q}}}) & =\int f^{\pm}\,\mathrm{d}^{3}p^{\pm}\\
 & =\frac{N^{\pm}e^{-a^{\pm}r^{2}-b^{\pm}G(r)-c^{\pm}K(r)}}{4\pi^{2}R\int_{0}^{+\infty}re^{-a^{\pm}r^{2}-b^{\pm}G(r)-c^{\pm}K(r)}\mathrm{d}r}\:,
\end{aligned}
\label{eq:-26}
\end{equation}
and the charge radial density
\begin{equation}
\begin{aligned}\rho^{\pm}(r) & =q^{\pm}\frac{\int n^{\pm}(\text{\ensuremath{\mathbf{q}}})r\,\mathrm{d}\theta\,\mathrm{d}z}{\int r\,\mathrm{d}\theta\,\mathrm{d}z}\\
 & =\frac{q^{\pm}}{\mathscr{V^{\pm}}}e^{-a^{\pm}r^{2}-b^{\pm}G(r)-c^{\pm}K(r)}
\end{aligned}
\label{eq:-29}
\end{equation}

\noindent with $\mathscr{V^{\pm}}=\frac{4\pi^{2}R\int_{0}^{+\infty}re^{-a^{\pm}r^{2}-b^{\pm}G(r)-c^{\pm}K(r)}\mathrm{d}r}{N^{\pm}}$.
In order to move to self consistency, we as well compute, by component,
the currents densities induced. Since

\begin{equation}
v_{z}^{\pm}=\frac{1}{m^{\pm}}\left(p_{z}^{\pm}-q^{\pm}B_{0}G(r)\right)
\end{equation}

\noindent and

\begin{equation}
v_{\theta}^{\pm}=\frac{1}{m^{\pm}}\left(\frac{p_{\theta}^{\pm}}{r}-q^{\pm}B_{0}\left(\frac{r}{2}+\frac{1}{r}K(r)\right)\right)\:,
\end{equation}

\noindent we obtain after integration $j_{\theta}^{\pm}(\text{\ensuremath{\mathbf{q}}})$
and $j_{z}^{\pm}(\text{\ensuremath{\mathbf{q}}})$, so the full current
densities are given by

\begin{equation}
\text{\ensuremath{\mathbf{J}}}^{\pm}(\text{r})=-\frac{1}{\beta^{\pm}}\left(\gamma_{\theta}^{\pm}r\,\mathbf{e}_{\theta}+\gamma_{z}^{\pm}\,\mathbf{e}_{z}\right)\rho^{\pm}(r)\:.\label{eq:-34-1}
\end{equation}

\noindent Furthermore, we point out the relations

\begin{equation}
\frac{j_{\theta}^{\pm}(\text{\ensuremath{\mathbf{q}}})}{rj_{z}^{\pm}(\text{\ensuremath{\mathbf{q}}})}=\frac{\gamma_{\theta}^{\pm}}{\gamma_{z}^{\pm}}\:.
\end{equation}
We now move to the full self-consistent equation, we remain in Coulomb
gauge ($\nabla\mathbf{.A}=0$), so we have

\begin{equation}
\Delta\mathbf{A}=-\mu_{0}\left(\mathbf{J^{+}}+\mathbf{J^{-}}\right)\label{eq:-35}
\end{equation}

\noindent and we end up with the self-consistent equation

\begin{equation}
\begin{cases}
\frac{1}{r}\frac{\partial}{\partial r}\left(\frac{1}{r}\frac{\partial}{\partial r}K(r)\right) & =\frac{\mu_{0}}{B_{0}}\left[\frac{\gamma_{\theta}^{+}}{\beta^{+}}\rho^{+}(r)+\frac{\gamma_{\theta}^{-}}{\beta^{-}}\rho^{-}(r)\right]\\
\frac{1}{r}\frac{\partial}{\partial r}\left(r\frac{\partial}{\partial r}G(r)\right) & =\frac{\mu_{0}}{B_{0}}\left[\frac{\gamma_{z}^{+}}{\beta^{+}}\rho^{+}(r)+\frac{\gamma_{z}^{-}}{\beta^{-}}\rho^{-}(r)\right]
\end{cases}\:.\label{eq:-39-2}
\end{equation}
We recall that we are assuming no electric field, so we have to impose
electro-neutrality

\begin{equation}
\rho^{+}(r)+\rho^{-}(r)=0
\end{equation}
that implies

\begin{equation}
\begin{cases}
\frac{1}{r}\frac{\partial}{\partial r}\left(\frac{1}{r}\frac{\partial}{\partial r}K(r)\right) & =\frac{\mu_{0}}{B_{0}}\left[\frac{\gamma_{\theta}^{+}}{\beta^{+}}-\frac{\gamma_{\theta}^{-}}{\beta^{-}}\right]\rho^{+}(r)\\
\frac{1}{r}\frac{\partial}{\partial r}\left(r\frac{\partial}{\partial r}G(r)\right) & =\frac{\mu_{0}}{B_{0}}\left[\frac{\gamma_{z}^{+}}{\beta^{+}}-\frac{\gamma_{z}^{-}}{\beta^{-}}\right]\rho^{+}(r)
\end{cases}\:.\label{eq:-39-1}
\end{equation}
We end up with a form of equations that are formally identical to
the ones found in the case of a single species with neutralizing background:

\begin{equation}
\begin{cases}
\frac{1}{r}\frac{\partial}{\partial r}\left(\frac{1}{r}\frac{\partial}{\partial r}K(r)\right) & =\kappa_{\theta}e^{-a^{+}r^{2}-b^{+}G(r)-c^{+}K(r)}\\
\frac{1}{r}\frac{\partial}{\partial r}\left(r\frac{\partial}{\partial r}G(r)\right) & =\kappa_{z}e^{-a^{+}r^{2}-b^{+}G(r)-c^{+}K(r)}
\end{cases}\:,
\end{equation}

\noindent where 
\begin{equation}
\kappa_{\theta/z}=\frac{\mu_{0}}{B_{0}}\left[\frac{\gamma_{\theta/z}^{+}}{\beta^{+}}-\frac{\gamma_{\theta/z}^{-}}{\beta^{-}}\right]\frac{q^{+}}{\mathscr{\mathscr{V^{+}}}}
\end{equation}

\noindent or with more details

\noindent {\footnotesize{}
\begin{equation}
\kappa_{\theta/z}=\frac{\mu_{0}}{B_{0}}\left[\frac{\gamma_{\theta/z}^{+}}{\beta^{+}}-\frac{\gamma_{\theta/z}^{-}}{\beta^{-}}\right]\frac{q^{+}N^{+}}{4\pi^{2}R\int_{0}^{+\infty}re^{-a^{+}r^{2}-b^{+}G(r)-c^{+}K(r)}\mathrm{d}r}\:.
\end{equation}
}\bibliographystyle{apsrev4-2}

\end{document}